\documentclass[runningheads]{cl2emult}

\usepackage{makeidx}  
\usepackage{graphicx} 
\usepackage{subeqnar} 
\usepackage{multicol} 
\usepackage{cropmark} 
\usepackage{math}     
\makeindex            

\begin{document}

\title*{Financial Time Series and Statistical Mechanics}

\toctitle{Financial Time Series and Statistical Mechanics}
\titlerunning{Financial Time Series and Statistical Mechanics}

\author{Marcel Ausloos }
\authorrunning{Marcel Ausloos }

\institute{GRASP\cite{byline} and SUPRAS\cite{byline}, \\ B5 Sart 
Tilman Campus, \\
B-4000, Li\`ege, Belgium}

\maketitle             

\begin{abstract}

A few characteristic exponents describing power law behaviors of roughness,
coherence and persistence in stochastic time series are compared to each other.
Relevant techniques for analyzing such time series are recalled in order to
distinguish how the various exponents are measured, and what basic differences
exist between each one. Financial time series, like the JPY/DEM and USD/DEM
exchange rates are
used for illustration, but mathematical ones, like (fractional or 
not) Brownian walks
can be used also as indicated.

\end{abstract}


\section{Introduction}

A great challenge in modern times is the construction of predictive 
theories for
nonlinear dynamical systems for which the evolution equations are barely known,
if known at all. General or so-called universal laws are aimed at 
from very noisy
data. A universal law should hold for different systems characterized by
different models, but leading to similar basic parameters, like the critical
exponents, depending only on the dimensionality of the system and the number of
components of the order parameter. This {\it in fine} leads to a 
predictive value
or power of the universal laws.

In order to obtain universal laws in stochastic systems one has to distinguish
true noise from chaotic behavior, and sort out coherent sequences from random
ones in experimentally obtained signals \cite{cellucci}. The stochastic aspects 
are not only
found in the statistical distribution of underlying frequencies characterizing
the Fourier transform of the signal, but also in the amplitude fluctuation
distribution and high moments or correlation functions.

Following the scaling hypothesis idea, neither time nor length scales 
have to be
considered \cite{refA1,addison,falconer}.  Henceforth the fractal geometry is a perfect framework for
studies of stochastic systems which do not appear at first to have underlying
scales.  A universal law can be a so-called {\it scaling} law if a $log[y(x)]$
vs. $log (x)$ plot gives a straight line (over several decades if possible)
leading to a slope measurement and the exponent characterizing the power law.

When examining such phenomena, it is often recognized that some {\it coherent
factor} is implied. Yet there are states which cannot be reached without going
through intricate evolutions, implying  concepts like {\it transience} and {\it
persistence}, - well known if one recalls the turbulence phenomenon 
and its basic
theoretical understanding \cite{turbulence}. Finally, the {\it apparent roughness} of the
signal can be put in mathematical terms. These concepts are briefly elaborated
upon in Sect. 2.

In the $y(x)$ function, $y$ and $x$ can be many ''things''. However 
the relevant
outlined concepts can be well illustrated when $x$ is the time $t$ variable. In
so doing time series serve as fundamental testing grounds. Several 
series can be
found in the literature. Financial time series and mathematical ones 
based on the
Weierstrass-Mandelbrot function \cite{berrylewis} describing fractional (or not) 
Brownian
walks can be used for illustration. The number of points should be large enough
to obtain small error bars. A few useful references, among many others,
discussing tests and other basic or technical considerations on non linear time
series analysis are to be found in \cite{book,rapp,schreiber,kantz,diks,malamud}.

Mathematical series, like (fractional or not) Brownian motions ($Bm$) and
practical ones, like financial time series, are thus of interest for discussion
or illustrations as done in Sect.3. There are several papers and books for
interest geared at financial time series analysis ... and 
forecasting. Again not
all can be mentioned, though see \cite{brockwell,franses,gourieroux}. 
On a more general basis, an
introduction to financial market analysis {\it per se}, can be found in 
\cite{Blake}.

Nevertheless it should be considered that any "scaling exponent"  should be
robust in a statistical sense with respect to small changes in the 
data or in the
data analysis technique. If this is so some physics considerations and modeling
can be pursued. One question is often raised for statistical purposes 
whether the
data is {\it stationary} or not, i.e. whether the analyzed raw 
signal, or any of
its combinations depend on the (time) origin of the series. This theoretical
question seems somewhat practically irrelevant in financial, 
meteorological, ...
sciences because the data is obviously {\it never} stationary. In fact, in such
new ''exotic applications of physics''  a restricted criterion for stationarity
is thought to be sufficient: if the data statistical mean and the
whatever-extracted-parameter do not change too much (up to some statistical
significance~\cite{book}) the data is called $quasi-stationary$. If so it can be next
offered for fundamental investigations. Thereafter, the prefix ''quasi'' is
immediately forgotten and not written anymore.

Several characteristics plots leading to {\it universality 
considerations}, thus
fractal-like exponents are first to be recalled. They are obtained 
from different
techniques which are briefly reviewed for completeness either in Sect. 3 or in
Appendices. Some technical materials can be usefully found in \cite{3a,3b,3c}.
 This should serve to distinguish how the various
exponents are measured, and what basic differences exist between each one. The
numerical values pertaining to the words (i) persistence, (ii) coherence, and
(iii) roughness will be given and related to each other. From a practical point
of view, as illustrated in the exercice session which was taking 
place after the
lecture, the cases of foreign exchange currency rates, i.e. DEM/USD and DEM/JPY
are used. They are shown in Fig.1 and 2 for a time interval ranging 
from Jan. 01,
1993 till June 30, 2000.

\begin{figure} \centering \includegraphics[width=.7\textwidth]{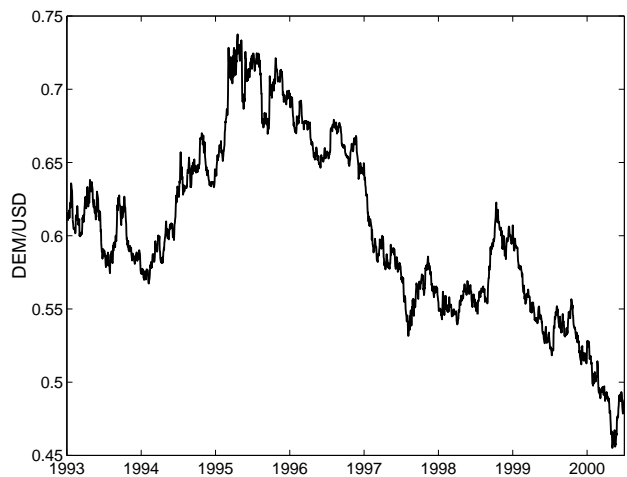}
\caption{Evolution of the exchange rate
DEM/USD from Jan. 01, 1993 till June 30, 2000} \label{eps1} \end{figure}

\begin{figure} \centering \includegraphics[width=.7\textwidth]{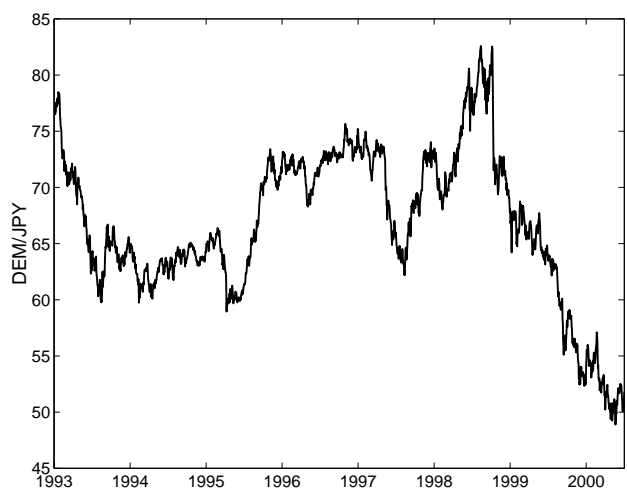}
\caption{Evolution of the exchange rate
DEM/JPY from Jan. 01, 1993 till June 30, 2000} \label{eps2} \end{figure}

For an adequate perspective, let it be recalled that such modern concepts of
statistical physics have been recently applied in analyzing time series outside
finance as well, like in particular those arising from biology
\cite{DNA,siwy}, medicine \cite{westmed},  meteorology \cite{ivabuda}, 
electronics \cite{gate},
image recognition \cite{image},  ... again without intending to list all 
references of
interest as should be done in a (longer) review paper. Many examples 
can also be
found in this book through contributions by world specialists of computer
simulations.

\section{Phase, amplitude and frequency revisited}

\subsection{Coherence}

When mentioning the word coherence to any student in physics, he/she is
immediately thinking about ''lasers'', - that is where the word has been most
striking in any scientific memory. It is recalled that the laser is a so
interesting instrument because all photons are emitted in {\it phase},  more
exactly the difference in phases between emitted waves is a constant in time.
Thus there is a so-called {\it coherence} in the light beam. They are 
other cases
in which phase coherence occurs, let it be recalled that light bugs 
are emitting
coherently~\cite{bug}, young girls in dormitories have their 
period in a
coherent way, driving conditions are best if some coherence is
imposed~\cite{bel}, crystals have a better shape and properties if they
are grown in a ''coherent way''; sand piles and stock markets seem also to have
coherent properties \cite{4,5}.

\subsection{Roughness}

No need to say that a wave is characterized by its {\it amplitude} 
which has also
some known importance in measurements indeed, be it often  the measure of an
intensity (the square of the amplitude). More interestingly one can define the
{\it roughness} of a profile by observing how the signal amplitude 
varies in time
(and space if necessary), in particular the correlation between the various
amplitude fluctuations.

\subsection{Persistence}

On the other hand, it can be easily shown that a periodic signal can be
decomposed into a series of $sin$ or $cos$ for which the frequencies are in
arithmetic order. Thus, the third ''parameter'' of the wave is its {\it
frequency}. Some (regular) frequency effect is surely apparent in all cycling
phenomena, starting from biology, climatology, meteorology, astronomy, but also
stock markets, foreign currency exchange markets, tectonics events, traffic and
turbulence, and politics. For non periodic signals, the Fourier transform has
been introduced in order to sort out the distribution of frequencies 
of interest,
i.e. the ''density of modes''. The  distribution defines the sort of {\it
persistence} of a phenomenon. It might be also examined whether the frequencies
are distributed  in a geometrical progression, rather than following an
ordinary/usual arithmetic progression, i.e. whether the phenomena might be
log-periodical, like in antennas, earthquakes and stock market crashes
\cite{sornettejpn,8}.

\section{Power Law Exponents}

\subsection{Persistence and Spectral Density }

Data from a (usually discrete) time series $y(t)$ are one dimensional sets and
are more simple to analyze at first than spatial ones
\cite{prakash,wesfreid}. Below and for simplicity one 
considers that the
measurements are taken at equal time intervals. Thus for financial time series,
there is no holiday nor week-end. A more general situation is hardly necessary
here. Two classic examples of a mathematical univariate stochastic time series
are those resulting from Brownian motion and Levy walk cases \cite{west}. In 
both cases,
the power spectral density $S_1(f)$ of the (supposed to be self-affine) time
series $y(t)$ has a single power-law dependence on the frequency $f$,

\begin{equation} S_1(f)  ~\sim ~ f^{-\beta}, \end{equation}

following from the Fourier transform

\begin{equation} S_1(f) =  ~\int ~ dt  ~e^{ift}  ~y(t). \end{equation}

For $y(t)$ one can use the Weierstrass-Mandelbrot (fractal) function 
\cite{berrylewis}
\begin{equation} {\cal W}(t) = \sum_{\rm m} ~ \gamma^{(2-D)m}\, [1 ~-
~e^{i\gamma^n t}] ~e^{i\phi_nt}, \end{equation}

with $\gamma>1$, and $1<D<2$. The phase $\phi_n$ can be stochastic or
deterministic. For illustration, Berry and Lewis \cite{berrylewis} 
took $\phi_n$ = n $\mu$,
with $\mu$ = 0 or $\pi$. The function obeys

\begin{equation} {\cal W}(\gamma t) = ~\gamma^{(2-D)}\, 
~e^{-i\gamma^\mu} ~{\cal
W}(t) , \end{equation} is stationary, and its trend, in the 
deterministic cases,
is given by

\begin{equation} {\cal W}\prime(t) \sim {\cal W}\prime\prime(t)  \sim t^{2-D} /
ln~\gamma. \end{equation}

The power spectrum is easily calculated \cite{berrylewis} to be
\begin{equation} S_1(f) \sim ...(1/ln~\gamma) ~f^{2D-5}. \end{equation}

One could search whether the function moments obey power laws with 
characterizing
exponents. Equation(1) allows one to put the phenomena into the {\it self-affine}
class of persistent phenomena characterized by the $\beta$ value. The 
range over
which $\beta$ is well defined in Eq.(1) indicates the range of the 
persistence in
the time series. A Brownian motion is characterized by $\beta$ = 2, and a white
noise by $\beta$ = 0.\footnote{\it Notice that the differences between adjacent
values of a Brownian motion amplitude result in white noise.} See a very
interesting set of such mathematical signals and the corresponding 
power spectrum in \cite{malamud}.

\begin{figure} \centering \includegraphics[width=.7\textwidth]{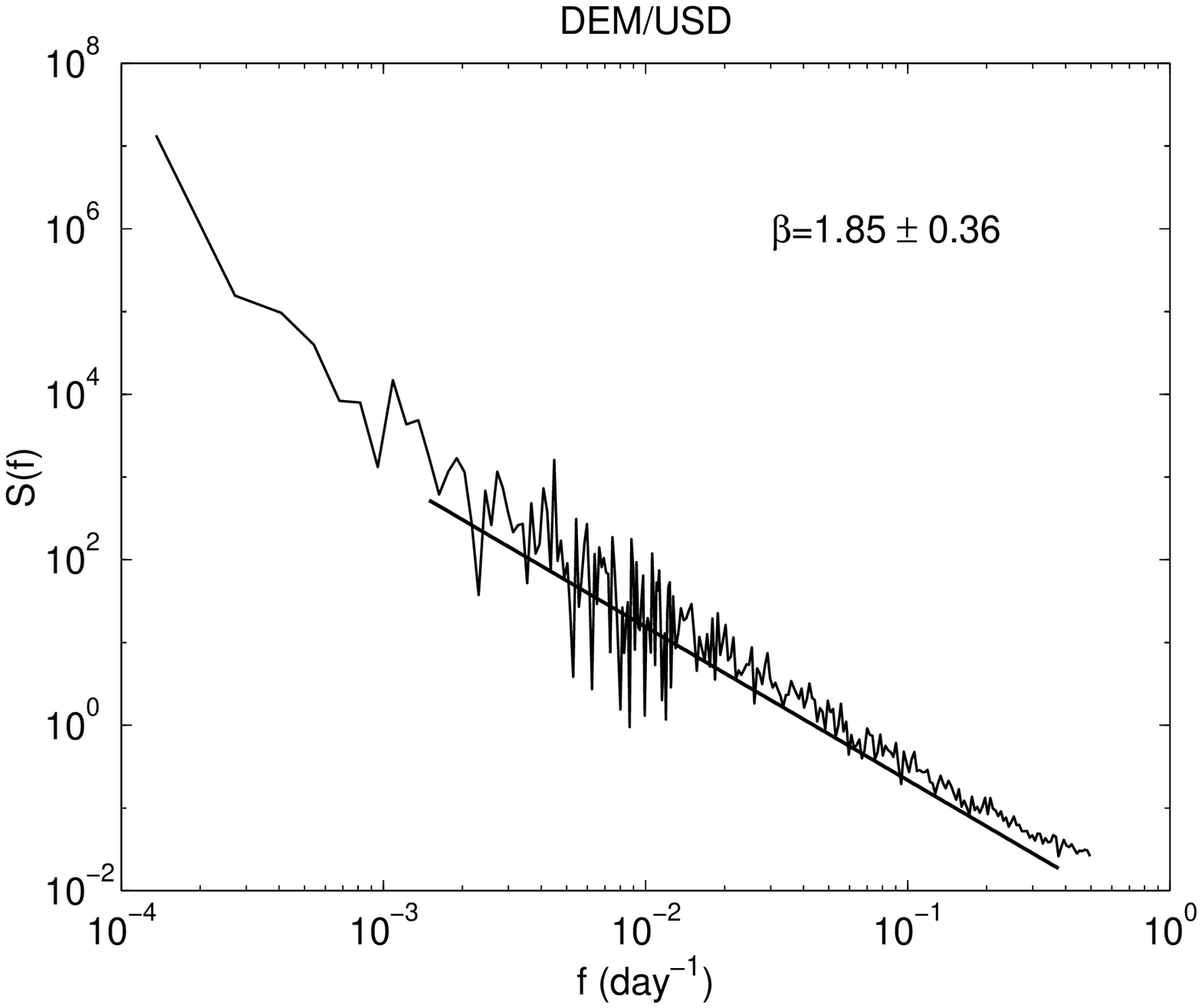}
\caption{The power spectrum of the DEM/USD exchange rate for
the time interval data in Fig.1} \label{eps3} \end{figure}

\begin{figure} \centering \includegraphics[width=.7\textwidth]{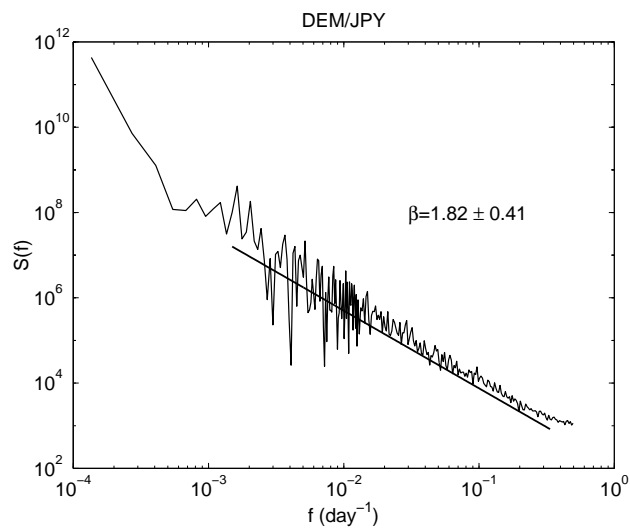}
\caption{The power spectrum of the DEM/JPY exchange rate for
the time interval data in Fig.2} \label{eps4} \end{figure}

The Fourier transform, or power spectrum,  of the financial signals used for
illustration here are found in Fig. 3 and Fig. 4. Notice the large error bars,
allowing to estimate that $\beta$ is about equal to 2, as for a 
trivial Brownian
motion case. However the coherence and/or roughness aspect are masked in this
one-shot analysis. Only the persistence behavior is touched upon.

If the distribution of fluctuations is not a power law, or if marked deviations
exist, say the {\it statistical correlation coefficient} is less than $0.99$ ,
indicating that a mere power law for $S(f)$ is doubtful, a more thorough search
of the basic frequencies is in order. A crucial step is to extract 
deterministic
or stochastic components~\cite{cellucci}, e.g. the stochastic aspects found in 
the statistical distribution of values show its persistence to be either 
nonexistent
(white noise case) or existent, i.e. $\beta \neq 2$. If so the 
persistence can be
qualitatively thought to be strong or weak. The Fourier transform can sometimes
indicate the presence of specific frequencies, much more abundant 
than others, in
particular if cycles exist, as in meteorology and climatology.

The range of the persistence is obtained from the correlations 
between events. A
''short'' or ''long range'' is checked through the autocorrelation function,
usually $c_1$. This function is a particular case of the so-called 
''$q-$th order
structure function'' \cite{Vicsek} or ''$q-$th order height-height
correlation function'' of the (normalized) time-dependent signal $y(t_i)$,

\begin{equation}
c_q(\tau)=\left<|y(t_{i+r})-y(t_i)|^q\right>_{\tau}/\left<|y(t_i)|^q\right>_{\tau}
\end{equation} where only non-zero terms are considered in the 
average ${\langle.
\rangle}_{\tau}$ taken over all couples $(t_{i+r},t_i)$ such that $\tau
=|t_{i+r}-t_i|$. In so doing one can obtain a set of exponents $\beta_q$

If the autocorrelation is larger than unity for some long time $t$ one can talk
about strong persistence, otherwise it is weak. This criterion 
defines the scale
of time $t$ in $y(t)$ for which there is long or short (time) range 
persistence.
Notice that the lower limit of the time scale is due to the 
discretization step,
and this sets the highest frequency to be the inverse of twice the 
discretizatin
interval. The upper limit is obvious.

\subsection{Roughness, Fractal Dimension, Hurst exponent \\and Detrended
Fluctuation Analysis }

The fractal dimensioncite{refA1,addison,falconer,roughness} $D$ is often used 
to characterize the roughness of
profiles \cite{roughness}.  Several methods are used for measuring $D$,
like the box counting method, not quite but are not quite efficient; 
many others
are found in the literature as seen in \cite{refA1,addison,falconer,west}
and here below. For topologically one
dimensional systems, the fractal dimension $D$ is related to the 
exponent $\beta$ by

\begin{equation} \beta = 5-2D. \end{equation}

A Brownian motion is characterized by $D$ = 3/2, and a white noise by $D$ = 2.0
\cite{berrylewis,malamud}.

Another ''measure'' of a signal roughness is sometimes given by the Hurst $Hu$
exponent, first defined in the ''rescale range theory'' (of
Hurst~\cite{Hu4,Hu5} ) who measured the Nile flooding and drought
amplitudes. The Hurst method consists in listing the differences between the
observed value at a discrete time $t$ over an interval with size $N$ 
on which the
mean has been taken. The upper ($y_M$) and lower ($y_m$) values in 
that interval
define the range $R_N = y_M - y_m$. The root mean square deviation $S_N$ being
also calculated, the ''rescaled range'' is $R_N/S_N$ is expected to behave like
$N^{Hu}$. This means that for a (discrete) self-affine signal $y(t)$, the
neighborhood of a particular point on the signal can be rescaled by a 
factor $b$
using the roughness (or Hurst \cite{addison,falconer}) 
exponent $Hu$ and
defining the new signal $ b^{-Hu} y(bt)$. For the exponent value $Hu$, the
frequency dependence of the signal so obtained should be undistinguishable from
the original one, i.e. $y(t)$.

The roughness (Hurst) exponent $Hu$ can be calculated from the height-height
correlation function $c_1(\tau)$ supposed to behave like \begin{equation}
c_1(\tau) = {\langle {| y(n_{i+r})-y(n_i) |} \rangle}_{\tau} \sim {\tau}^{H_1}
\end{equation} whereas \begin{equation} Hu = 1 +  H_1, \end{equation} 
rather than
from the box counting method. For a {\it persistent} signal, 
$H_1~>~1/2$; for an
{\it anti-persistent} signal, $H_1~<~1/2$. Flandrin has theoretically proved
\cite{Flandrin} that

\begin{equation} \beta = 2 Hu - 1, \end{equation} thus $\beta$ = 1+2 
$H_1$. This
implies that the  classical random walk (Brownian motion) is such 
that $Hu=3/2$.
It is clear that

\begin{equation} D = 3 - Hu. \end{equation}

Fractional Brownian motion values are practically found to lie between $1$ and
$2$ \cite{3a,3b,4}.  Since a white noise is a truly random process, it
can be concluded that $Hu~ =~1.5$  implies an uncorrelated time series
\cite{west}.

Thus $D>1.5$, or $Hu<1.5$ implies antipersistence and $D<1.5$, or $Hu>1.5$
implies persistence. From preimposed $Hu$ values of a fractional 
Brownian motion
series, it is found that the equality here above usually holds true in a very
limited range and $\beta$ only slowly converges toward the value $Hu$
\cite{malamud}.

The inertia axes of the {\it 2-variability diagram}
\cite{kiandma1,bablo,kiandma2} seem to be related to these values and
could be used for fast measurements as well.

The above results can be compared to those obtained from the Detrended
Fluctuation Analysis ~ \cite{4,ndub} ($DFA$) method. $DFA$ \cite{DNA} consists
in dividing a random variable sequence $y(n)$ over $N$ points into $N/\tau$
boxes, each containing $\tau$ points. The best linear trend $z(n)=an+b$ in each
box is defined. The fluctuation function $F(\tau)$ is then calculated following
\begin{equation} {F^2(\tau)} = {1 \over \tau} {\sum_{n=(k-1)\tau+1}^{k\tau}
{|y(n)-z(n)|}^2}, {\hskip 1cm} k=1,2, \cdots, N/\tau. \end{equation} Averaging
$F(\tau)^2$ over the $N/\tau$ intervals gives the fluctuations 
$\langle F(\tau)^2
\rangle$ as a function of $\tau$. If the $y(n)$ data are random uncorrelated
variables or short range correlated variables, the behavior is expected to be a
power law \begin{equation} \langle F^2 \rangle ^{1/2} \sim \tau^{Ha}
\end{equation} with $Ha$ different from $0.5$.

The exponent $Ha$ is so-labelled for Hausdorff 
\cite{addison,west}. It is
expected, not always proved as emphasized by
\cite{berrylewis,mandelbrot} that

\begin{equation} Ha = 2 - D , \end{equation} where $D$ is the 
self-affine fractal
dimension \cite{refA1,west}. It is immediately seen that 
\begin{equation}
\beta = 1 + 2 Ha. \end{equation}

For Brownian motion, $Ha=0.5$, while for white noise $Ha=0$ and $D=2$.

\begin{figure} \centering \includegraphics[width=.7\textwidth]{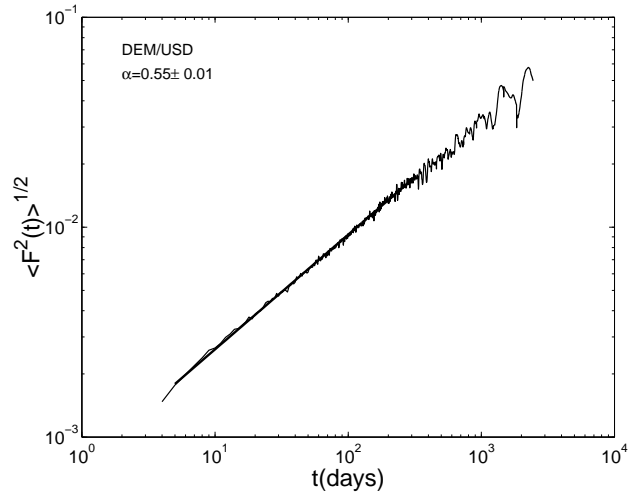}
\caption{The DFA result for the DEM/USD exchange rate for
the time interval data in Fig.1} \label{eps5} \end{figure}

\begin{figure} \centering \includegraphics[width=.7\textwidth]{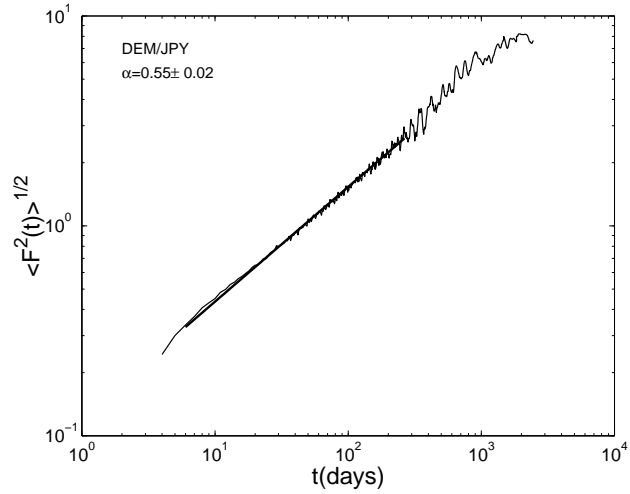}
\caption{The DFA result for the DEM/JPY exchange rate for
the time interval data in Fig.2} \label{eps6} \end{figure}

The $DFA$ log-log plots of the DEM/JPY and DEM/USD exchange rates are given in
Fig. 5 and Fig. 6. It is seen that the value of $H_a$ fulfills the above
relations for the DEM/JPY and DEM/USD data, since $H_a$ = 0.55. Both 
$Ha$ and $D$
readily measure the roughness and persistence strength. Fractional Brownian
motion $Ha$ values are found to lie between $0$ and $1$. The effect 
of a trend is
supposedly eliminated here.  However only the linear or cubic detrending have
been studied to my knowledge \cite{ndub}.  Other trends, 
emphasizing some
characteristic frequency, like seasonal cycles, could be further studied.

A generalized Hurst exponent $H(q)$ is defined through the relation
\begin{equation} c_q(\tau) \propto {\tau}^{qH(q)}, \quad q\ge 0  \end{equation}
where $c_q(\tau)$ has been defined here above

The {\it intermittency} of the signal can be studied through the so-called
singular measure analysis of the small-scale gradient field obtained from the
data through

\begin{equation} \varepsilon(r;l)=\frac{r^{-1}\sum_{i=l}^{l+r-1}
|y(t_{i+r})-y(t_i)|} {<|y(t_{i+r})-y(t_i)|>} \end{equation} with 
\begin{equation}
\qquad i=0, \dots , \Lambda - r \end{equation} and 
\begin{equation}\qquad r=1, 2,
\dots , \Lambda = 2^m \qquad ,\end{equation} where $m$ is an integer. 
The scaling
properties of the generating function are then searched for through 
the equation

\begin{equation} \chi_q(\tau)=<\varepsilon(r;l)^q>\sim \tau^{-K(q)} ,\quad q\ge
0, \end{equation} with $\tau$ as defined above.

The $K(q)$-exponent is closely related to the generalized dimensions
$D_q=1-K(q)/(q-1)$ \cite{HP83}. The nonlinearity of both 
characteristics
exponents, $qH(q)$ and $K(q)$, describes the multifractality of the 
signal. If a
linear dependence is obtained, then the signal is monofractal or in 
other words,
the data follows a simple scaling law for these values of $q$. Thus 
the exponent
\cite{Vicsek,Davis}  \begin{equation} C_1 = \left. {dK_q \over dq}
\right|_{q=1} \end{equation}  is a measure of the intermittency lying in the
signal $y(n)$ and can be numerically estimated by measuring $K_q$ around $q=1$.
Some conjecture on the role/meaning of $H_1$ is found  in \cite{4}.
 From some financial and political data analysis it seems that $H_1$ 
is a measure
of the information entropy of the system.

\section{Conclusion}

It has been emphasized that to analyze stochastic time series, like those
describing fractional Brownian motion and foreign currency exchange 
rates reduces
to examining the distribution of and correlations between amplitudes, 
frequencies
an phases of harmonic-like components of the signal. Due to some scaling
hypothesis, characteristic exponents can be obtained to describe 
power laws. The
usefulness of such exponents serves in determining universality 
classes, and {\it
in fine} building physical or algorithmic models. In the case of 
financial times
series, the exponents can even serve into imagining some investment strategy
\cite{4}.  Notice that due to the non stationarity of the data, such
exponents vary with time, and multifractal concepts must be brought in at a
refining stage, - including in an investment strategy.

In that spirit, let it be emphasized here the analogy between a 
$H_1,C_1$ diagram
and the $\omega,k$ diagram of dynamical second order phase transitions
\cite{HES}. In the latter the frequency and the phase of a time signal
are considered on the same footing, and encompass the critical and 
hydrodynamical
regions. In the present cases an analog diagram relates the roughness and
intermittency. This has been already examined in \cite{MW69,BR94}.

Among other various physical data analysis techniques which have been recently
presented in order to obtain some information on the deterministic 
and/or chaotic
content of univariate data, let us point out the wavelet technique 
which has been
considerably used (several references exist, see below) including for DNA and
meteorology studies \cite{Davis,Bacry,Arneodo,struzik}. The $H_1,C_1$ technique 
used in turbulence and
meteorology \cite{kiackerm} is also somewhat appropriate.

\begin{table}[ht] \begin{center} \caption{Values of the most relevant exponents
in various regimes (i.e., stationary, persistent, antipersistent) of univariate
stochastic series : $D$ : fractal dimension; $Ha$ : Hausdorff measure; $Hu$ :
Hurst exponent ; $\alpha$ : from DFA technique; $\beta$ : power spectrum
exponent: $WN$ = white noise, $(f)Bm$ : (fractional) Brownian motion; $flat$ :
flat spectrum.}

\begin{tabular}{|c|c|c|c|c|c|c|} Signal name & {$D $} & {$Ha $} & {$Hu $} &
{$\alpha $} & {$\beta $} & {$ $} \\  
\hline 
$- $ & $- $ & $- $ & $0 $ 
& $- $ & $-1
$ & $- $ \\

$ - $ & $ - $ & $ - $ & $ - $ & $ antipersist $ & $ - $ & $ station $ \\

$-$ & $-$ & $-$ & $0.5$ & $0.5$ & $0$ & $uncorrel$ \\

$-$ & $-$ & $-$ & $-$ & $persist$ & $-$ & $station$ \\

$WN$ & $2$ & $0$ & $1$ & $-$ & $1$ & $-$ \\

$fBm$ & $-$ & $-$ & $-$ & $superpersist$ & $-$ & $nonstat$ \\

$Bm$ & $3/2$ & $0.5$ & $3/2$ & $-$ & $2$ & $-$ \\

$flat$ & $1$ & $1$ & $2$ & $superpersist$ & $3$ & $-$ \\ \hline \end{tabular}

\end{center} \end{table}

There are many other techniques which are not mentioned here, like the weighted
fixed point \cite{yukalov}, and the time-delay embedding
\cite{cao}. Surely several others have been used, but only 
those relevant
for the present purpose have been fully mentioned hereabove. As a 
summary of the
above, Table 1 indicates the range of values found for different signals and
their relationship to stationarity, persistence and coherence. In the 
many years
to come, it seems relevant to ask for more data on multivariate functions, thus
extending the above considerations to other real mathematical and 
physical cases
in higher dimensions.

Finally, two short Appendices should follow in order to remain consistent with
the oral lectures. It was shown that another time series ''analysis'' technique
is often used by experts for some predictability purpose, i.e. the {\it moving
average technique}. It is briefly discussed in Appendix A. It has served in the
{\it \"{u}bungen}. Also, these lecture notes would be incomplete without
mentioning the intrinsic {\it discrete scale invariance} implication in time
series. Such a substructure leads to log-periodic oscillations in the time
series, whence to fascinating effects and surprises in predicting crash-like
events.  This is mentioned in Appendix B.

\vskip 1.0cm {\noindent \large \bf Acknowledgements}

Thanks to K. Ivanova for very useful discussions, and for preparing 
the data and
drawings. A.  P\c{e}kalski kindly reviewed the manuscript and has much improved
it due to his usual and well known pedagogical insight. Great thanks go to the
organizers of the WE-Heraeus-Ferienkurse f\"{u}r Physik in den Neuen
Bundesl\"{a}ndern 2000, Chemnitz, "Vom Billardtisch bis Monte Carlo - 
Spielfelder
der Statistischen Physik" for inviting me to present these ideas and 
techniques,
and to the students for their questions, comments and output during the {\it
\"{u}bungen}.

\section{Appendix A : moving averages} To take into account the trend can be
shown to be irrelevant for such short range correlation events. The trend is
anyway quite ill defined since it is a statistical mean, and thus 
depends on the
size of the interval, i.e. the number of data points which is taken 
into account.
Nevertheless many technical analyses rely on signal averages over various time
intervals, like the moving average method \cite{movaverage}. These
methods shold be examined form a physical point of view. An interesting
observation has resulted from checking the density of intersections 
of such mean
values over different time interval windows, which are continuously 
shifted. This
corresponds to obtaining a spectrum of the so-called moving averages
\cite{movaverage}, used by analysts in order to point to ''gold'' or
''death'' crosses in a market. The density $\rho$ of crossing points 
between any
two moving averages is obviously a measure of long-range power-law correlations
in the signal. It has been found that $\rho$ is a symmetric function of $\Delta
T$, i.e. the difference between the interval sizes on which the averages are
taken, and it has a simple power law form \cite{mav1,mav2}. 
This leads to
a very fast and rather reliable measure of the fractal dimension of th signal.
The method can be easily implemented for obtaining the time evolution of $D$,
thus for elementary investment strategies.

\section{Appendix B : discrete scale invariance}

It has been proposed that an economic index $y(t)$ follows a complex power law
\cite{bouchaud,feigenbaum}, i.e.

\begin{equation} y(t) ~=~ A ~+ ~B~ (t_c-t)^{-m} ~[1 ~+ ~C ~cos(\omega
~ln((t_c-t)/t_c)~ + ~\phi)] \end{equation}

for $t<t_c$, where $t_c$ is the crash-time or rupture point, A, B, $m$, C,
$\omega$, $\phi$ are parameters. This index evolution is a power law ($m$)
divergence (for $m>0$) on which log-periodic ($\omega$) oscillations are taking
place. The law for $y(t)$ diverges at $t$=$t_c$ with an exponent $m$ 
(for $m>0$)
while the period of the oscillations converges to the rupture point at $t=t_c$.
This law is similar to that of critical points at so-called second order phase
transitions \cite{HES}, but generalizes the scaleless 
situation for cases
in which discrete scale invariance is presupposed 
\cite{sorphysrep}. This
relationship was already proposed in order to fit experimental measurements of
sound wave rate emissions prior to the rupture of heterogeneous composite
stressed up to failure \cite{souillard}. The same type of complex power
law behavior has been observed as a precursor of the Kobe earthquake in Japan
\cite{sornette}.

Fits using Eq.(23) were performed on the $S\&P500$ data
\cite{bouchaud,feigenbaum} for the period preceding the 1987 October
crash. The parameter values have not been found to be robust against small
perturbations, like a change in the phase of the signal. It is known 
in fact that
a nonlinear seven parameter fit is highly unstable from a numerical point of
view. Indeed, eliminating the contribution of the oscillations in Eq.(1), i.e.
setting C=0, implies that the best fit leads to an exponent $m=0.7$ 
quite larger
than $m=0.33$ for C $\neq$ 0 \cite{bouchaud}. Feigenbaum and Freund 
\cite{feigenbaum} also reported
various values of $m$ ranging from 0.53 to 0.06 for various indexes and events
(upsurges and crashes).

Universality in this case means that the value of $m$ should be the 
same for any
crash and for any index. In so doing a single model should describe the phase
transition, and the exponent would define the model and be the only 
parameter. A
limiting case of a power law behavior is the logarithmic behavior, 
corresponding
to $m$ = 0, i.e. the divergence of the index $y$ for $t$ close to 
$t_c$ should be

\begin{equation} y(t)~ = ~A~ + ~B ~ln((t_c-t)/t_c) ~[1 ~+~ C~ cos(\omega
ln((t_c-t)/t_c )~+~ \phi)]  \end{equation}

This logarithmic behavior is known in physics as characterizing the 
specific heat
("four point correlation function") of the Ising model, and the
Kosterlitz-Thouless phase transition \cite{kosterlitz} in spatial
dimensions equal to two. They are thus specific to systems with a low order
dimension of the order parameter. It is nevertheless a smooth transition. The
mean value of the order parameter \cite{HES} is not defined over long
range scales, but a phase transition nevertheless exists because there is some
ordered state on small scales. In addition to the physical 
interpretation of the
latter relationship, the advantages are that (i) the number of parameters is
reduced by one, and (ii) the log-divergence seems to be close to reality.

In order to test the validity of Eq.(24) in the vicinity of crashes, we have
separated the problems of the divergence itself and the oscillation 
convergences
on the other hand, in order to extract two values for the rupture point $t_c$:
(i) $t_{cdiv}$ for the power (or logarithmic) divergence and (ii) 
$t_{cosc}$ for
the oscillation convergence. In so doing the long range and short range
fluctuation scales are examined on an equal footing. The final $t_c$ 
is obtained
at the intersection of two straight lines, by successive iteration fits.  The
results of the fit as well as the correlation fitting factor $R$ have 
been given
in \cite{d1,d2}.

A technical point is in order : The rupture point $t_{cosc}$ is estimated by
selecting the maxima and the minima of the oscillations through a 
double envelope
technique \cite{d2}. Finally notice that the oscillation 
basic frequency
depends on the connectivity of the underlying space \cite{3b,3c}. The log-periodic
behavior also corresponds to a complex fractal dimension
\cite{sorphysrep,moussa}.

\end{document}